\documentclass[10pt,twocolumn]{article}
\usepackage[top=1.8cm, bottom=2cm, left=1.8cm, right=1.8cm]{geometry}
\usepackage{times}
\usepackage{booktabs} %
\usepackage{paralist}
\usepackage[small,bf]{caption}
\usepackage[keeplastbox]{flushend}
\usepackage{subfigure}
\usepackage{graphicx}

\usepackage[hyphens]{url}
\makeatletter
\def\url@leostyle{%
  \@ifundefined{selectfont}{\def\UrlFont{\small}}%
  {\def\UrlFont{}}%
}
\makeatother
\usepackage[hang,flushmargin]{footmisc}

\renewenvironment{thebibliography}[1]{
  \begin{oldthebibliography}{#1}
    \setlength{\itemsep}{0.1em}
    \setlength{\parskip}{0.1em}
}
{
  \end{oldthebibliography}
}

\renewcommand{\footnoterule}{%
  \kern -4pt
  \hrule width 1in 
  \kern 2pt
}

\usepackage[compact]{titlesec}
\titlespacing*{\section}{0pt}{*3}{3pt} 
\titlespacing{\subsection}{0pt}{*2}{2pt}

\newif\ifcomment
\commenttrue 
\ifcomment
\newcommand{\XXX}[2]{{\bf \textcolor{blue}{#1: #2}}}
\newcommand{\jbnote}[1]{{\bf \textcolor{magenta}{JB: #1}}}
\newcommand{\msnote}[1]{{\bf \textcolor{magenta}{MS: #1}}}
\newcommand{\gs}[1]{{\bf \textcolor{red}{gs: #1}}}
\newcommand{\edc}[1]{{\bf \textcolor{blue}{EDC: #1}}}
\else
\newcommand{\XXX}[2]{}
\newcommand{\jbnote}[1]{}
\newcommand{\msnote}[1]{}
\newcommand{\gs}[1]{}
\newcommand{\edc}[1]{}
\fi
\newcommand{\descr}[1]{\vspace{0.1cm}\noindent\textbf{#1}}

\begin{document}
\title{\bf What is Gab? A Bastion of Free Speech or\\ an Alt-Right Echo Chamber?}
\author{Savvas Zannettou$^{\star}$,  Barry Bradlyn$^\#$,  Emiliano De Cristofaro$^\dagger$,\\ Haewoon Kwak$^+$, Michael Sirivianos$^{\star}$, Gianluca Stringhini$^\dagger$, Jeremy Blackburn$^\ddagger$\\[0.5ex]
\normalsize $^{\star}$Cyprus University of Technology, ${^\#}$Princeton Center for Theoretical Science,$^\dagger$University College London, \\[-0.5ex]
\normalsize ${^+}$Qatar Computing Research Institute \& Hamad Bin Khalifa University, ${^\ddagger}$University of Alabama at Birmingham\\
\normalsize sa.zannettou@edu.cut.ac.cy, bbradlyn@princeton.edu, \{e.decristofaro,g.stringhini\}@ucl.ac.uk\\[-0.5ex]
\normalsize hkwak@hbku.edu.qa, michael.sirivianos@cut.ac.cy, blackburn@uab.edu}%

\date{}

\maketitle

\begin{abstract}
Over the past few years, a number of new ``fringe'' communities, like 4chan or certain subreddits, have gained traction on the Web at a rapid pace.
However, more often than not, little is known about how they evolve or what kind of activities they attract, despite recent research has shown that they influence how false information reaches mainstream communities. 
This motivates the need to monitor these communities and analyze their impact on the Web's information ecosystem.

In August 2016, a new social network called Gab was created as an alternative to Twitter. It positions itself as putting ``people and free speech first'', welcoming users banned or suspended from other social networks.
In this paper, we provide, to the best of our knowledge, the first characterization of Gab.
We collect and analyze 22M posts produced by 336K users between August 2016 and January 2018, finding that Gab is predominantly used for the dissemination and discussion of news and world events, and that it attracts alt-right users, conspiracy theorists, and other trolls.
We also measure the prevalence of hate speech on the platform, finding it to be much higher than Twitter, but lower than 4chan's Politically Incorrect board.
\end{abstract}

\section{Introduction}

The Web's information ecosystem is composed of multiple communities with varying influence~\cite{zannettou2017web}.
As mainstream online social networks become less novel, users have begun to join smaller, more focused platforms.
In particular, as the former have begun to reject fringe communities identified with racist and aggressive behavior, a number of alt-right focused services have been created.
Among these emerging communities, the Gab social network has attracted the interest of a large number of users since its creation in 2016~\cite{gab_site}, a few months before the US Presidential Election.
Gab was created, ostensibly as a censorship-free platform, aiming to protect free speech above anything else.
From the very beginning, site operators have welcomed users banned or suspended from platforms like Twitter for violating terms of service, often for abusive and/or hateful behavior.
In fact, there is extensive anecdotal evidence that the platform has become the alt-right's new hub~\cite{gab_alt_right} and that it exhibits a high volume of hate speech~\cite{gab_hate_speech} and racism~\cite{gab_racism}.
As a result, in 2017, both Google and Apple rejected Gab's mobile apps from their stores because of hate speech~\cite{gab_hate_speech} and non-compliance to pornographic content guidelines~\cite{gab_apple_porn}.

In this paper, we provide, to the best of our knowledge, the first characterization of the Gab social network.
We crawl the Gab platform and acquire 22M posts by 336K users over a 1.5 year period (August 2016 to January 2018).
Overall, the main findings of our analysis include:
\begin{enumerate}
\itemsep0em 
\item Gab attracts a wide variety of users, ranging from well-known alt-right personalities like Milo Yiannopoulos to conspiracy theorists like Alex Jones. We also find a number of ``troll'' accounts that have migrated over from other platforms like 4chan, or that have been heavily inspired by them.
\item Gab is predominantly used for the dissemination and discussion of world events, news, as well as conspiracy theories. Interestingly, we note that Gab reacts strongly to events related to white nationalism and Donald Trump.
\item Hate speech is extensively present on the platform, as we find that 5.4\% of the posts include hate words. This is 2.4 times higher than on Twitter, but 2.2 times lower than on 4chan's Politically Incorrect board (/pol/)~\cite{hine2017kek}.
\item There are several accounts making coordinated efforts towards recruiting millennials to the alt-right.
\end{enumerate}

In summary, our analysis highlights that Gab appears to be positioned at the border of mainstream social networks like Twitter and ``fringe'' Web communities like 4chan's /pol/. 
We find that, while Gab claims to be all about free speech, this seems to be merely a shield behind which its alt-right users hide.

\descr{Paper Organization.} In the next section, we review the related work. Then, in Section~\ref{sec:gab}, we provide an overview of the Gab platform, while in Section~\ref{sec:analysis} we present our analysis on Gab's user base and the content that gets shared.
Finally, the paper concludes in Section~\ref{sec:conclusion}.

\section{Related Work}
In this section, we review previous work on social network analysis and in particular on fringe communities.

Kwak et al.~\cite{kwak2010what} are among the first to study Twitter, aiming to understand its role on the Web.
They show that Twitter is a powerful network that can be exploited to assess human behavior on the Web.
However, the Web's information ecosystem does not naturally build on a single or a few Web communities;
with this motivation in mind, Zannettou et al.~\cite{zannettou2017web} study how mainstream and alternative news propagate across multiple Web communities, measuring the influence that each community have on each other.
Using a statistical model known as Hawkes Processes, they highlight that small ``fringe'' Web communities within Reddit and 4chan can have a substantial impact on large mainstream Web communities like Twitter.

With the same multi-platform point of view, Chandrasekharan et al.~\cite{chandrasekharan2017bag} propose an approach, called Bag of Communities, which aims to identify abusive content within a community. 
Using training data from nine communities within 4chan, Reddit, Voat, and Metafilter, they outperform approaches that focus only on in-community data.

Other work also focuses on characterizing relatively small alt-right Web communities.
Specifically, Hine et al.~\cite{hine2017kek} study 4chan's Politically Incorrect board (/pol/), and show that it attracts a high volume of hate speech.
They also find evidence of organized campaigns, called {\em raids}, that aim to disrupt the regular operation of other Web communities on the Web;
e.g., they show how 4chan users raid YouTube videos by posting large numbers of abusive comments in a relatively small period of time.

Overall, our work is, to the best of our knowledge, the first to study the Gab social network, analyzing what kind of users it attracts, what are the main topics of discussions, and to what extent Gab users share hateful content.

\begin{table*}[t]
\centering
\resizebox{0.99\textwidth}{!}{
\begin{tabular}{@{}lllllllll@{}}
\toprule
\multicolumn{3}{c}{\textbf{Followers}}                                      & \multicolumn{3}{c}{\textbf{Scores}}        & \multicolumn{3}{c}{\textbf{PageRank}}                        \\ \midrule
\textbf{Name}             & \textbf{Username} & \textbf{\#}                 & \textbf{Name}            & \textbf{Username}      & \textbf{\#}  & \textbf{Name}            & \textbf{Username}      & \textbf{PR score} \\ \midrule
Milo Yiannopoulos         & m                 & \multicolumn{1}{r|}{45,060} & Andrew Torba             & a                      &  \multicolumn{1}{r|}{819,363}    & Milo Yiannopoulos & m & 0.013655\\
PrisonPlanet              & PrisonPlanet      & \multicolumn{1}{r|}{45,059} & John Rivers              & JohnRivers             & \multicolumn{1}{r|}{606,623}     & Andrew Torba  & a & 0.012818\\
Andrew Torba              & a                 & \multicolumn{1}{r|}{38,101} & Ricky Vaughn             & Ricky\_Vaughn99        & \multicolumn{1}{r|}{496,962}   & PrisonPlanet & PrisonPlanet & 0.011762  \\
Ricky Vaughn              & Ricky\_Vaughn99   & \multicolumn{1}{r|}{30,870} & Don                      & Don                    & \multicolumn{1}{r|}{368,698 }    &Mike Cernovich & Cernovich & 0.006549\\
Mike Cernovich            & Cernovich         & \multicolumn{1}{r|}{29,081} & Jared Wyand              & JaredWyand             & \multicolumn{1}{r|}{281,798 }   & Ricky Vaughn  &Ricky\_Vaughn99 & 0.006143\\
Stefan Molyneux           & stefanmolyneux    & \multicolumn{1}{r|}{26,337} & $[omitted]$              & TukkRivers             & \multicolumn{1}{r|}{253,781}    &Sargon of Akkad & Sargonofakkad100 & 0.005823 \\
Brittany Pettibone        & BrittPettibone    & \multicolumn{1}{r|}{24,799} & Brittany Pettibone       & BrittPettibone         & \multicolumn{1}{r|}{244,025}    & $[omitted$] & d\_seaman & 0.005104 \\
Jebs                      & DeadNotSleeping   & \multicolumn{1}{r|}{22,659} & Tony Jackson             & USMC-Devildog          & \multicolumn{1}{r|}{228,370}     & Stefan Molyneux & stefanmolyneux & 0.004830\\
$[omitted]$ & TexasYankee4      & \multicolumn{1}{r|}{20,079} & [omitted]               & causticbob             & \multicolumn{1}{r|}{228,316}     & Brittany Pettibone & BrittPettibone & 0.004218\\
$[omitted]$     & RightSmarts       & \multicolumn{1}{l|}{20,042} & Constitutional Drunk     & USSANews               & \multicolumn{1}{r|}{224,261}     & Vox Day & voxday & 0.003972\\
Vox Day                   & voxday            & \multicolumn{1}{l|}{19,454} & Truth Whisper            & truthwhisper           & \multicolumn{1}{r|}{206,516}     &Alex Jones & RealAlexJones & 0.003345\\
$[omitted]$              & d\_seaman         & \multicolumn{1}{l|}{18,080} & Andrew Anglin            & AndrewAnglin           & \multicolumn{1}{r|}{203,437}  &Lauren Southern & LaurenSouthern & 0.002984  \\
Alex Jones                & RealAlexJones     & \multicolumn{1}{l|}{17,613} & Kek\_Magician            & Kek\_Magician          & \multicolumn{1}{r|}{193,819}     &Donald J Trump & realdonaldtrump & 0.002895\\
Jared Wyand               & JaredWyand        & \multicolumn{1}{l|}{16,975} & $[omitted]$           & shorty                 & \multicolumn{1}{r|}{169,167 }   &Dave Cullen &DaveCullen & 0.002824 \\
Ann Coulter               & AnnCoulter        & \multicolumn{1}{l|}{16,605} & $[omitted]$ & SergeiDimitrovicIvanov & \multicolumn{1}{r|}{169,091} & $[omitted]$ & e & 0.002648   \\
Lift                      & lift              & \multicolumn{1}{l|}{16,544} & Kolja Bonke              & KoljaBonke             & \multicolumn{1}{r|}{160,246}     & Chuck C Johnson & Chuckcjohnson & 0.002599\\
Survivor Medic            & SurvivorMed       & \multicolumn{1}{l|}{16,382} & Party On Weimerica       & CuckShamer             & \multicolumn{1}{r|}{155,021}  & Andrew Anglin & AndrewAnglin & 0.002599   \\
$[omitted]$          & SalguodNos        & \multicolumn{1}{l|}{16,124} & PrisonPlanet             & PrisonPlanet           & \multicolumn{1}{r|}{154,829}     & Jared Wyand & JaredWyand & 0.002504\\
Proud Deplorable          & luther            & \multicolumn{1}{l|}{15,036} & Vox Day                  & voxday                 & \multicolumn{1}{r|}{150,930}     & Pax Dickinson & pax & 0.002400\\
Lauren Southern           & LaurenSouthern    & \multicolumn{1}{l|}{14,827} & W.O. Cassity             & wocassity              & \multicolumn{1}{r|}{144,875}    &Baked Alaska &apple & 0.002292\\ \bottomrule
\end{tabular}
}
\caption{Top 20 popular users on Gab according to the number of followers, their score, and their ranking based on PageRank in the followers/followings network. We omit the ``screen names'' of certain accounts for ethical reasons.}
\label{tbl:top_20_users}
\end{table*}

\begin{figure*}[t]
\subfigure[]{\includegraphics[width=0.32\textwidth]{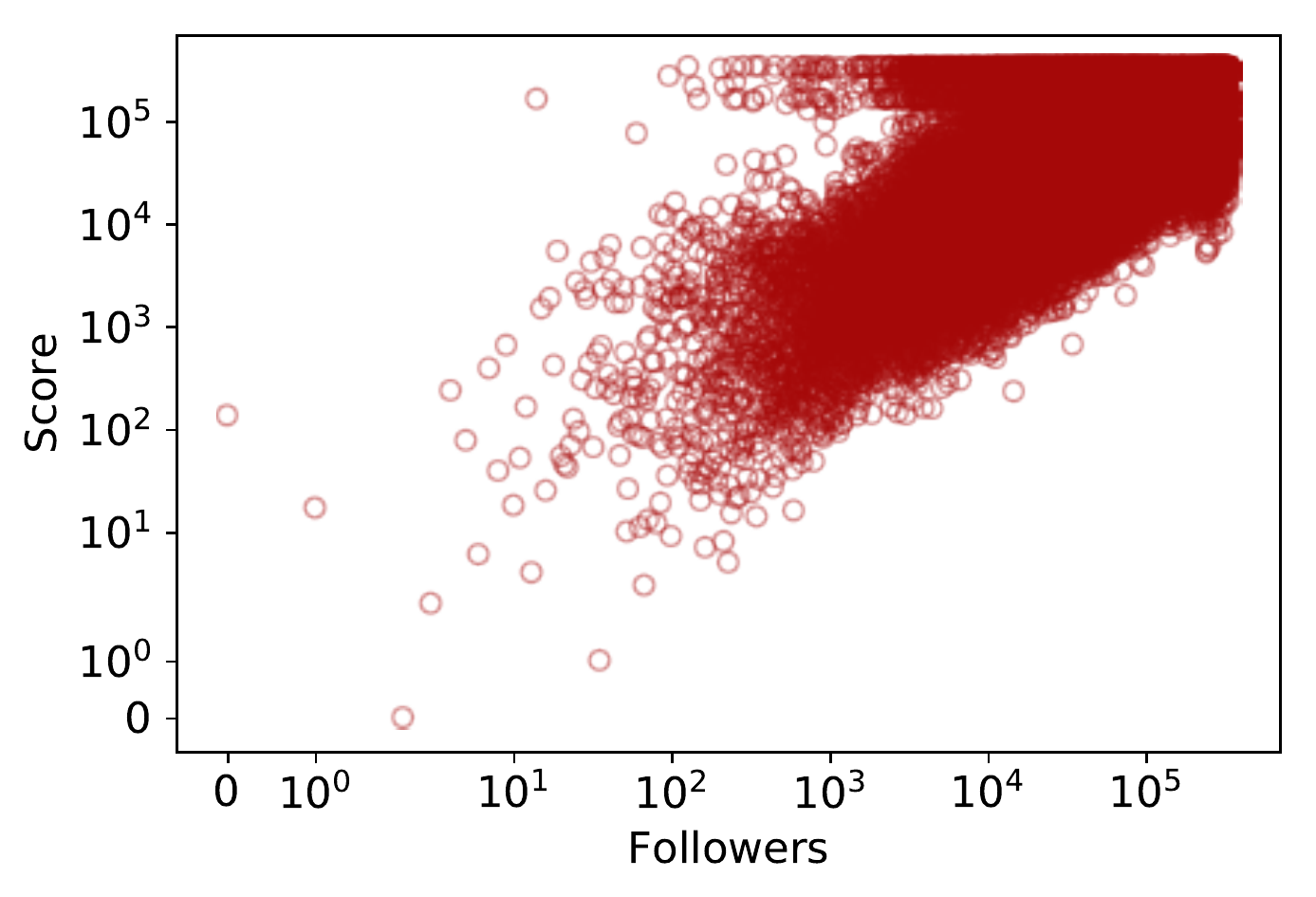}\label{subfig:followers_score}}
\subfigure[]{\includegraphics[width=0.32\textwidth]{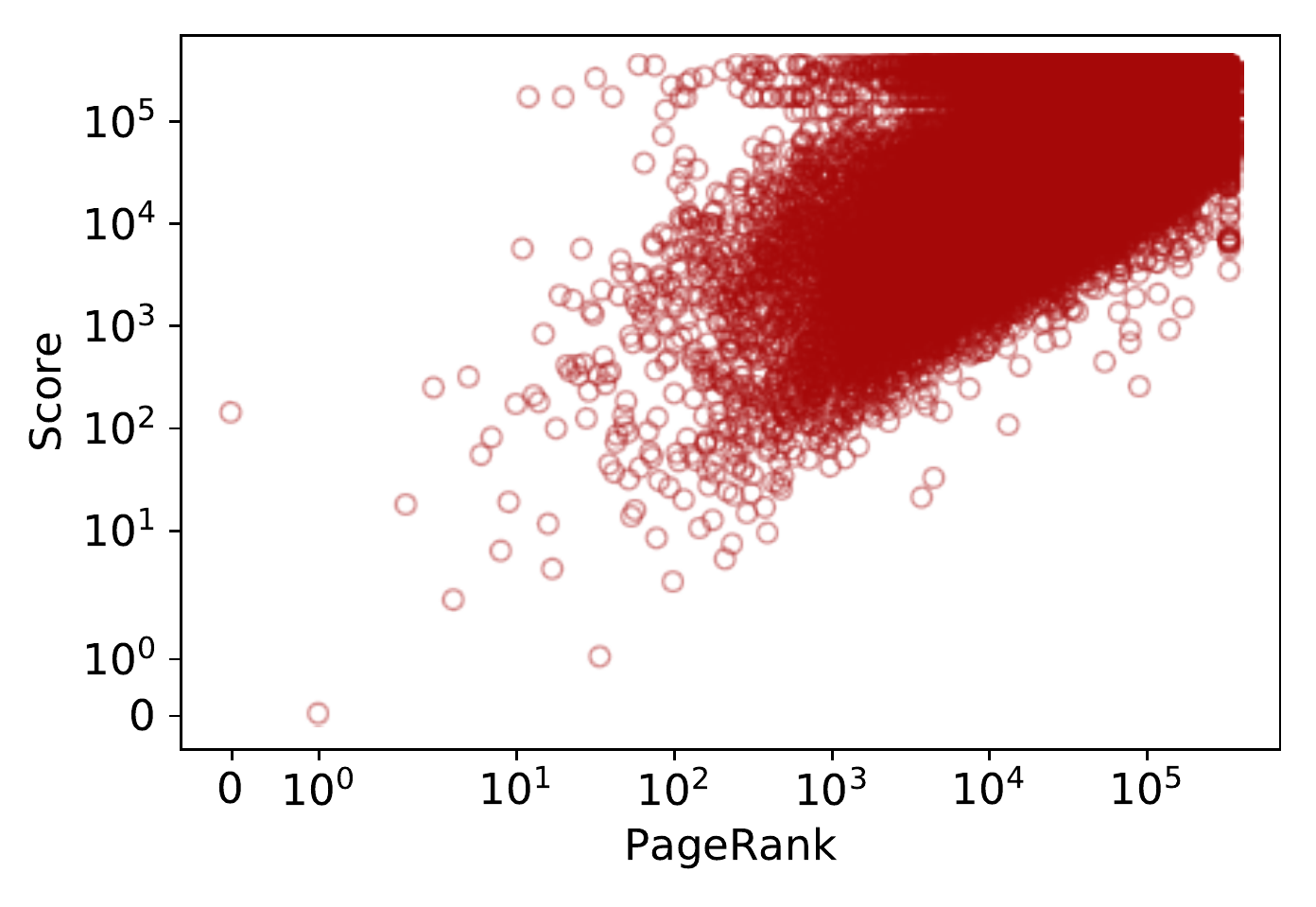}\label{subfig:pr_score}}
\subfigure[]{\includegraphics[width=0.32\textwidth]{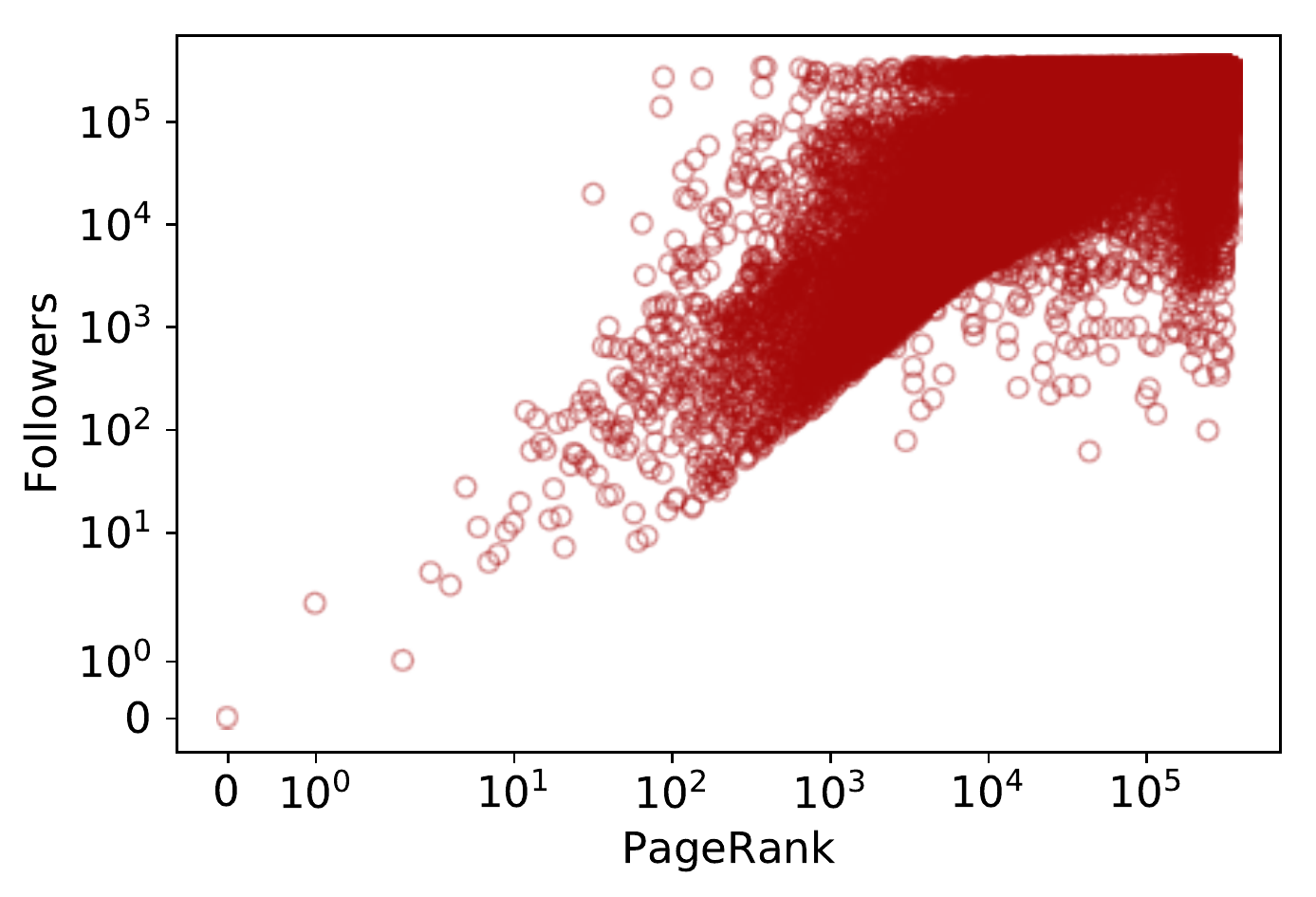}\label{subfig:pr_followers}}
\caption{Correlation of the rankings for each pair of rankings: (a) Followers - Score; (b) PageRank - Score; and (c) PageRank - Followers.}
\label{fig:correlations_scatter}
\end{figure*}

\section{Gab}\label{sec:gab}
Gab is a new social network, launched in August 2016, that ``champions free speech, individual liberty, and the free flow of information online.\footnote{\url{http://gab.ai}}''
It combines social networking features that exist in popular social platforms like Reddit and Twitter.
A user can broadcast 300-character messages, called ``gabs,'' to their followers (akin to Twitter).
From Reddit, Gab takes a modified voting system (which we discuss later).
Gab allows the posting of pornographic and obscene content, as long as users label it as Not-Safe-For-Work (NSFW).\footnote{What constitutes NSFW material is not well defined.}
Posts can be reposted, quoted, and used as replies to other gabs. 
Similar to Twitter, Gab supports hashtags, which allow indexing and querying for gabs, as well as mentions, which allow users to refer to other users in their gabs.

\descr{Topics and Categories.} Gab posts can be assigned to a specific \emph{topic} or \emph{category}.
\emph{Topics} focus on a particular event or timely topic of discussion and can be created by Gab users themselves; all topics are publicly available and other users can post gabs related to topics.
\emph{Categories} on the other hand, are defined by Gab itself, with 15 categories defined at the time of this writing.
Note that assigning a gab to a category and/or topic is \emph{optional}, and Gab moderates topics, removing any that do not comply with the platform's guidelines.

\descr{Voting system.} Gab posts can get up- and down-voted; a feature that determines the popularity of the content in the platform (akin to Reddit).
Additionally, each user has its own score, which is the sum of up-votes minus the sum of down-votes that it received to all his posts (similar to Reddit's user karma score~\cite{reddit_karma}).
This user-level score determines the popularity of the user and is used in a way unique to Gab: a user must have a score of at least 250 points to be able to down-vote other users' content, and every time a user down-votes a post a point from his user-level score is deducted.
In other words, a user's score is used as a form of currency expended to down-vote content.

\descr{Moderation.} Gab has a lax moderation policy that allows most things to be posted, with a few exceptions.
Specifically, it only forbids posts that contain ``illegal pornography'' (legal pornography is permitted), posts that promote terrorist acts, threats to other users, and doxing other users' personal information~\cite{snyder2017fifteen}.\footnote{For more information on Gab's guidelines, see \url{https://gab.ai/about/guidelines}.}

\descr{Monetization.} Gab is ad-free and relies on direct user support.
On October 4, 2016 Gab's CEO Andrew Torba announced that users were able to donate to Gab~\cite{gab_donate}.
Later, Gab added ``pro'' accounts as well.
``Pro'' users pay a per-month fee granting additional features like live-stream broadcasts, account verification, extended character count (up to 3K characters per gab), special formatting in posts (e.g., italics, bold, etc.), as well as premium content creation.
The latter allows users to create ``premium'' content that can only be seen by subscribers of the user, which are users that pay a monthly fee to the content creator to be able to view his posts.
The premium content model allows for crowdfunding particular Gab users, similar to the way that Twitch and Patreon work.
Finally, Gab is in the process of raising money through an Initial Coin Offering (ICO) with the goal to offer a ``censorship-proof'' peer-to-peer social network that developers can build application on top~\cite{gab_ico}.

\descr{Dataset.} Using Gab's API, we crawl the social network using a snowball methodology. 
Specifically, we obtain data for the most popular users as returned by Gab's API and iteratively collect data from all their followers as well as their followings.
We collect three types of information: 1)~basic details about Gab accounts, including username, score, date of account creation; 2)~all the posts for each Gab user in our dataset; and 3)~all the followers and followings of each user that allow us to build the following/followers network.
Overall, we collect 22,112,812 posts from 336,752 users, between August 2016 and January 2018.

\section{Analysis} \label{sec:analysis}
In this section, we provide our analysis on the Gab platform. 
Specifically, we analyze Gab's user base and posts that get shared across several axes.

\subsection{Ranking of users}
To get a better handle on the interests of Gab users, we first examine the most popular users using three metrics: 1)~the number of followers; 2)~user account score; and 3)~user PageRank.
These three metrics provide us a good overview of things in terms of ``reach,'' appreciation of content production, and importance in terms of position within the social network.
We report the top 20 users for each metric in Table~\ref{tbl:top_20_users}.
Although we believe that their existence in Table~\ref{tbl:top_20_users} is arguably indicative of their public figure status, for ethical reasons, we omit the ``screen names'' for accounts in cases where a potential link between the screen name and the user's real life names existed \emph{and} it was unclear to us whether or not the user is a public figure.
While Twitter has many celebrities in the most popular users~\cite{kwak2010what}, Gab seems to have what can at best be described as alt-right celebrities like Milo Yiannopoulos and Mike Cernovich.
\begin{table}[t]
\centering
\small
\begin{tabular}{@{}lrll@{}}
\textbf{Word} & \multicolumn{1}{l}{\textbf{(\%)}} & \textbf{Bigram} & \textbf{(\%)}              \\ \midrule
maga          & \multicolumn{1}{r|}{4.35\%}       & free speech     & \multicolumn{1}{r}{1.24\%} \\
twitter       & \multicolumn{1}{r|}{3.62\%}       & trump supporter & \multicolumn{1}{r}{0.74\%} \\
trump         & \multicolumn{1}{r|}{3.53\%}       & night area      & \multicolumn{1}{r}{0.49\%} \\
conservative  & \multicolumn{1}{r|}{3.47\%}       & area wanna      & \multicolumn{1}{r}{0.48\%} \\
free          & \multicolumn{1}{r|}{3.08\%}       & husband father  & \multicolumn{1}{r}{0.45\%} \\
love          & \multicolumn{1}{r|}{3.03\%}       & check link      & \multicolumn{1}{r}{0.42\%} \\
people        & \multicolumn{1}{r|}{2.76\%}       & freedom speech  & \multicolumn{1}{r}{0.41\%} \\
life          & \multicolumn{1}{r|}{2.70\%}       & hey guys        & \multicolumn{1}{r}{0.40\%} \\
like          & \multicolumn{1}{r|}{2.67\%}       & donald trump    & \multicolumn{1}{r}{0.40\%} \\
man           & \multicolumn{1}{r|}{2.49\%}       & man right       & 0.39\%                     \\
truth         & \multicolumn{1}{r|}{2.46\%}       & america great   & 0.39\%                     \\
god           & \multicolumn{1}{r|}{2.45\%}       & link contracts  & 0.35\%                     \\
world         & \multicolumn{1}{r|}{2.44\%}       & wanna check     & 0.34\%                     \\
freedom           & \multicolumn{1}{r|}{2.29\%}       & make america    & 0.34\%                     \\
right       & \multicolumn{1}{r|}{2.27\%}       & need man        & 0.34\%                     \\
american         & \multicolumn{1}{r|}{2.25\%}       & guys need       & 0.33\%                     \\
want      & \multicolumn{1}{r|}{2.23\%}       & president trump & 0.32\%                     \\
one          & \multicolumn{1}{r|}{2.20\%}       & guy sex         & 0.31\%                     \\
christian           & \multicolumn{1}{r|}{2.17\%}       & click link      & 0.30\%                     \\
time     & \multicolumn{1}{r|}{2.14\%}       & link login      & 0.30\%                     \\ \bottomrule
\end{tabular}
\caption{Top 20 words and bigrams found in the descriptions of Gab users.}
\label{tbl:users_description}
\end{table}

\descr{Number of followers.} The number of followers that each account has can be regarded as a metric of impact on the platform, as a user with many followers can share its posts to a large number of other users.
We observe a wide variety of different users; 1)~popular alt-right users like Milo Yiannopoulos, Mike Cernovich, Stefan Molyneux, and Brittany Pettibone; 2)~Gab's founder Andrew Torba; and 3)~popular conspiracy theorists like Alex Jones.
Notably lacking are users we might consider as counter-points to the alt-right right, an indication of Gab's heavily right-skewed user-base.

\descr{Score.} The score of each account is a metric of content popularity, as it determines the number of up-votes and down-votes that they receive from other users.
In other words, is the degree of appreciation from other users.
By looking at the ranking using the score, we observe two new additional categories of users: 1)~users purporting to be news outlets, likely pushing false or controversial information on the network like PrisonPlanet and USSANews; and 2)~troll users that seem to have migrated from or been inspired by other platforms (e.g., 4chan) like Kek\_Magician and CuckShamer.

\descr{PageRank.} We also compute PageRank on the followers/followings network and we rank the users according to the obtained score.
We use this metric as it quantifies the structural importance of nodes within a network according to its connections.
Here, we observe some interesting differences from the other two rankings.
For example, the account with username ``realdonaldtrump,'' an account reserved for Donald Trump, appears in the top users mainly because of the extremely high number of users that follow this account, despite the fact that it has no posts or score.

\begin{figure}[t]
\includegraphics[width=\columnwidth]{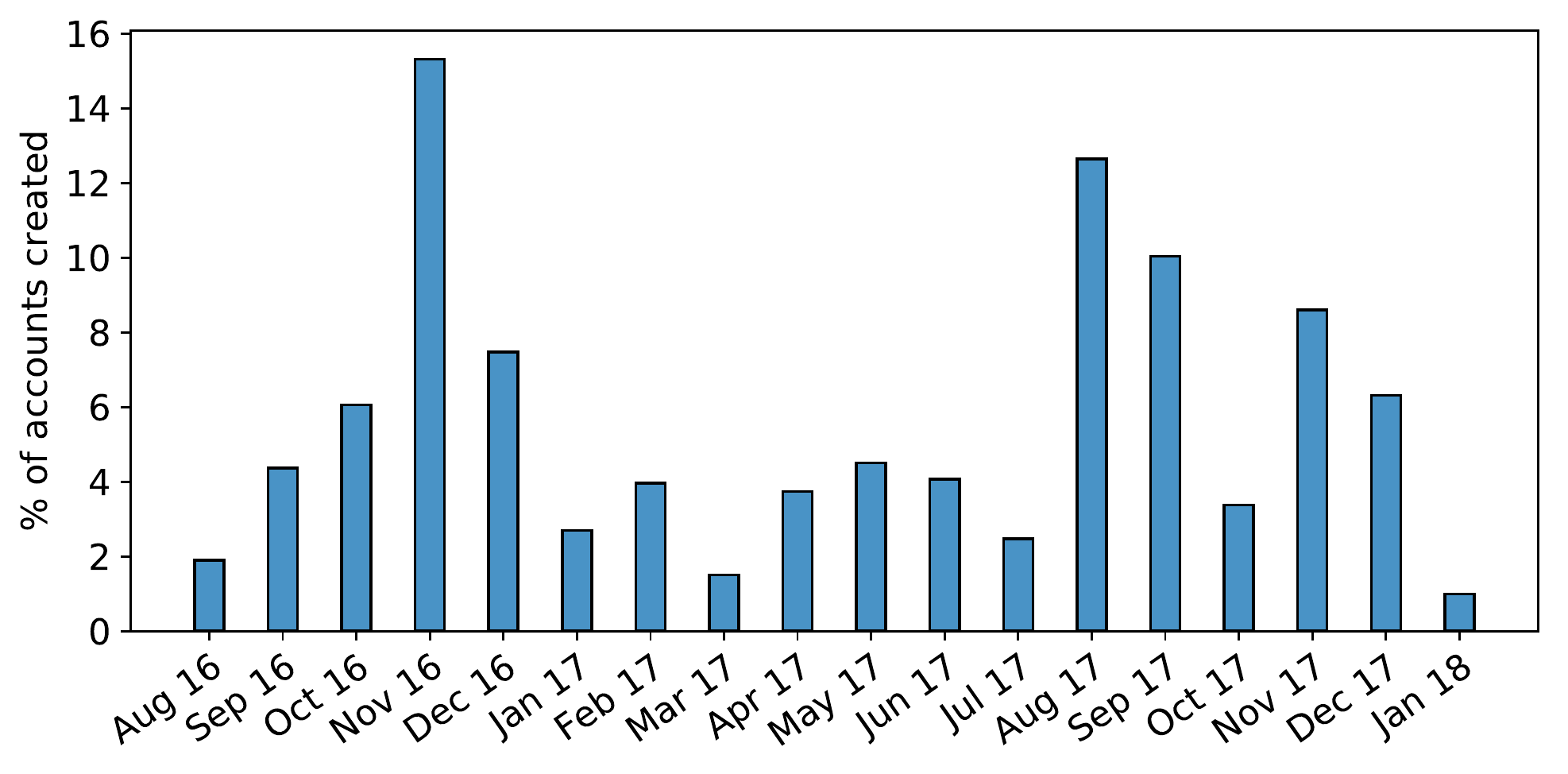}
\caption{Percentage of accounts created per month.}
\label{fig:bc_account_created}
\end{figure}

\descr{Comparison of rankings.} To compare the three aforementioned rankings, we plot the ranking of all the users for each pair of rankings in Fig.~\ref{fig:correlations_scatter}.
We observe that the pair with the most agreement is PageRank-Followers (Fig.~\ref{subfig:pr_followers}), followed by the pair Followers-Score (Fig,~\ref{subfig:followers_score}), while the pair with the least agreement is PageRank - Score (Fig~\ref{subfig:pr_score}.
Overall, for all pairs we find a varying degree of rank correlation.
Specifically, we calculate the Spearman's correlation coefficient for each pair of rankings; finding 0.53, 0.42, 0.26 for PageRank-Followers, Followers-Score, and PageRank-Score, respectively.
While these correlations are not terribly strong, they are significant ($p < 0.01$) for the two general classes of users: those that play an important structural role in the network, perhaps encouraging the diffusion of information, and those that produce content the community finds valuable.

\subsection{User account analysis}

\descr{User descriptions.} To further assess the type of users that the platform attracts we analyze the description of each created account in our dataset.
Note that by default Gab adds a quote from a famous person as the description of each account and a user can later change it.
Although not perfect, we look for any user description enclosed in quotes with a ``--'' followed by a name, and assume it is a default quote.
Using this heuristic, we find that only 20\% of the users actively change their description from the default.
Table~\ref{tbl:users_description} reports the top words and bigrams found in customized descriptions (we remove stop words for more meaningful results).
Examining the list, it is apparent that Gab users are conservative Americans, religious, and supporters of Donald Trump and ``free speech.''
We also find some accounts that are likely bots and trying to deceive users with their descriptions; among the top bigrams there some that nudge users to click on URLs, possibly malicious, with the promise that they will get sex. 
For example, we find many descriptions similar to the following: ``\emph{Do you wanna get sex tonight? One step is left ! Click the link - $<url>$}.''
It is also worth noting that our account (created for crawling the platform) was followed by 12 suspected bot accounts between December 2017 and January 2018 without making any interactions with the platform (i.e., our account has never made a post or followed any user).

\descr{User account creation.} We also look when users joined the Gab platform. Fig.~\ref{fig:bc_account_created} reports the percentage of accounts created for each month of our dataset.
Interestingly, we observe that we have peaks for account creation on November 2016 and August 2017. 
These findings highlight the fact that Gab became popular during notable world and politics events like the 2016 US elections as well as the Charlottesville Unite the Right rally~\cite{charlotesville}.
Finally, only a small percentage of Gab's users are either pro or verified, 0.75\% and 0.5\%, respectively, while 1.7\% of the users have a private account (i.e., only their followers can see their gabs).

\begin{figure*}[t!]
\center
\subfigure[]{\includegraphics[width=0.32\textwidth]{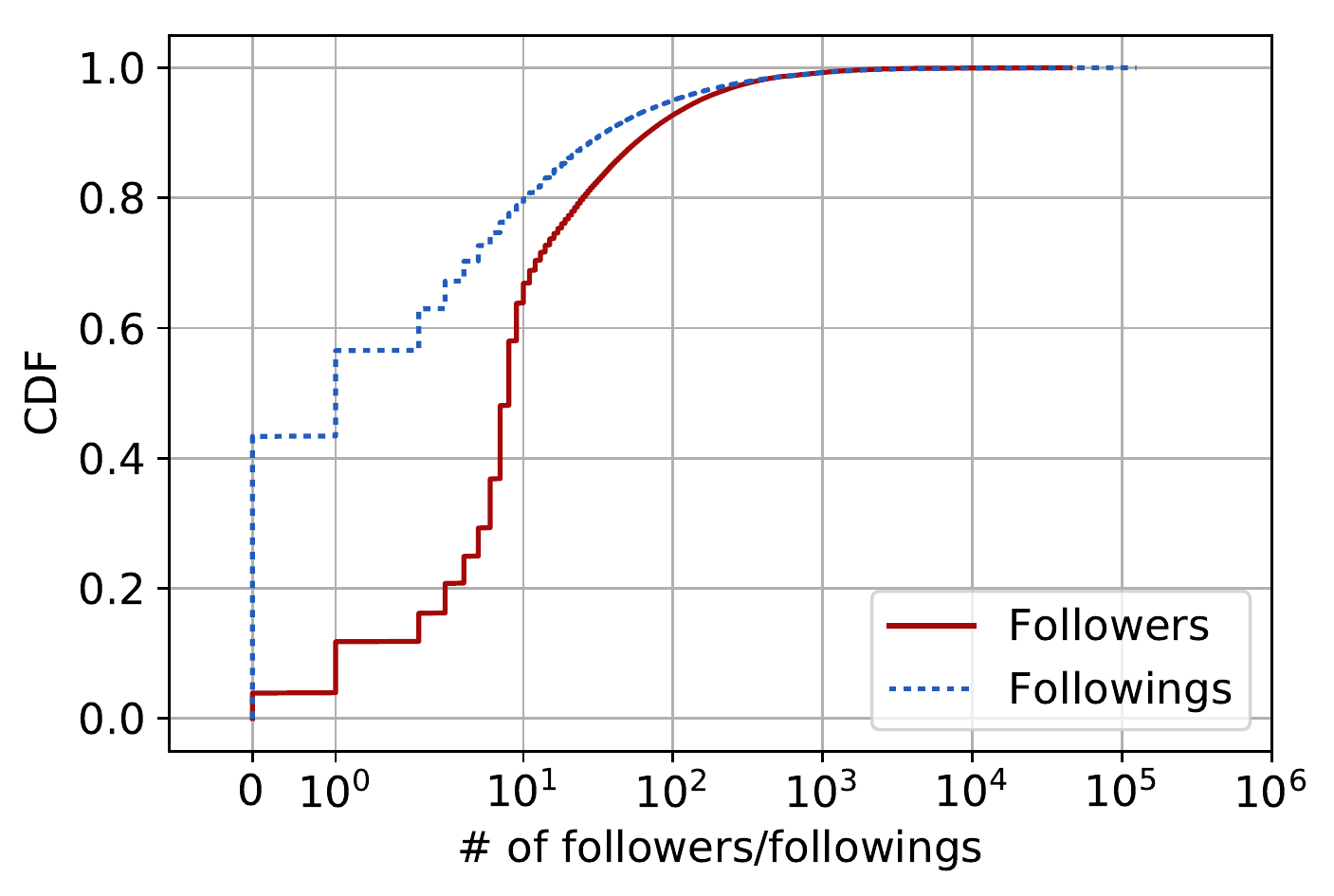}\label{fig:cdf_followers_following}}
\subfigure[]{\includegraphics[width=0.32\textwidth]{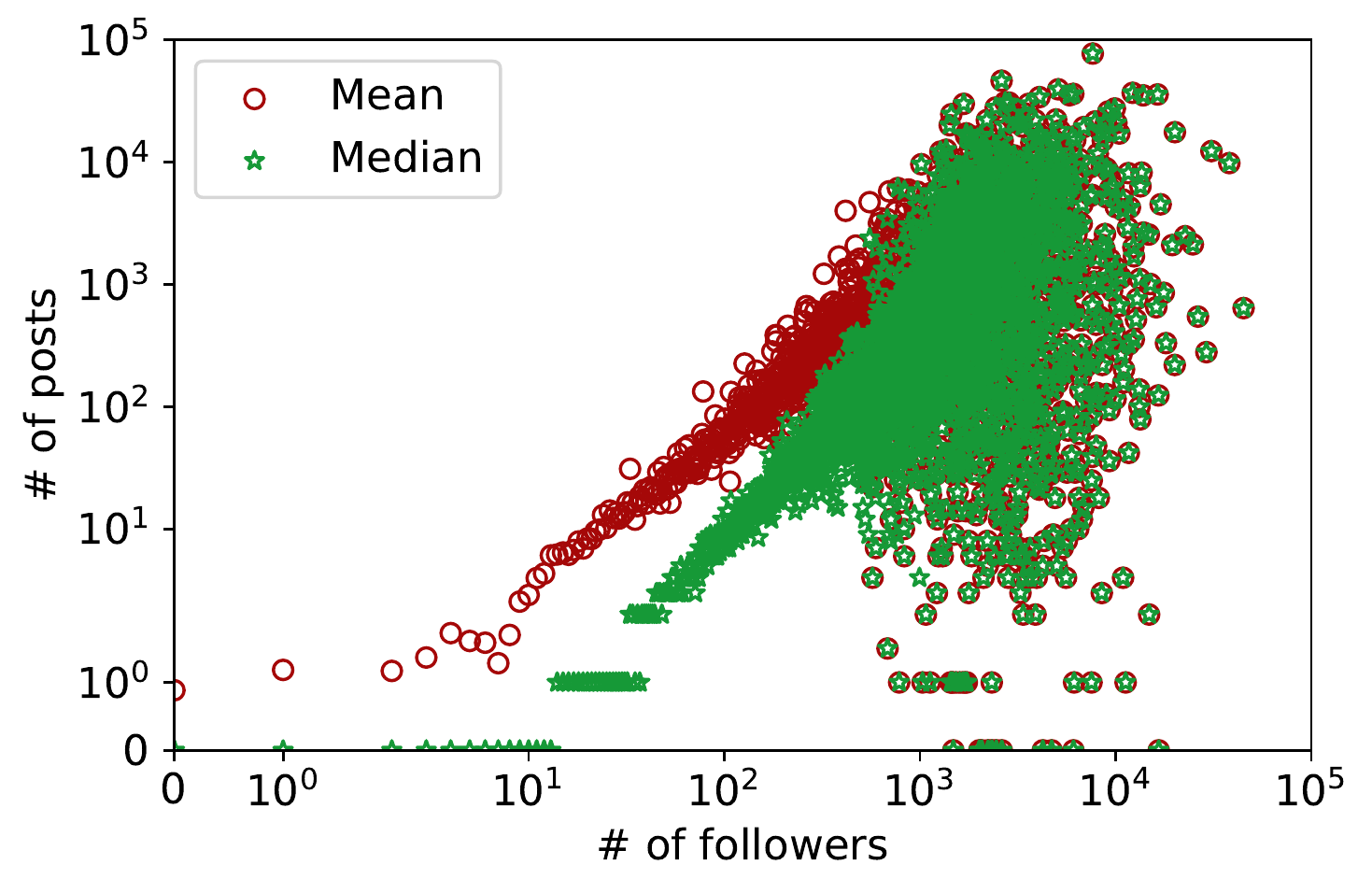}\label{subfig:scatter_followers_posts}}
\subfigure[]{\includegraphics[width=0.32\textwidth]{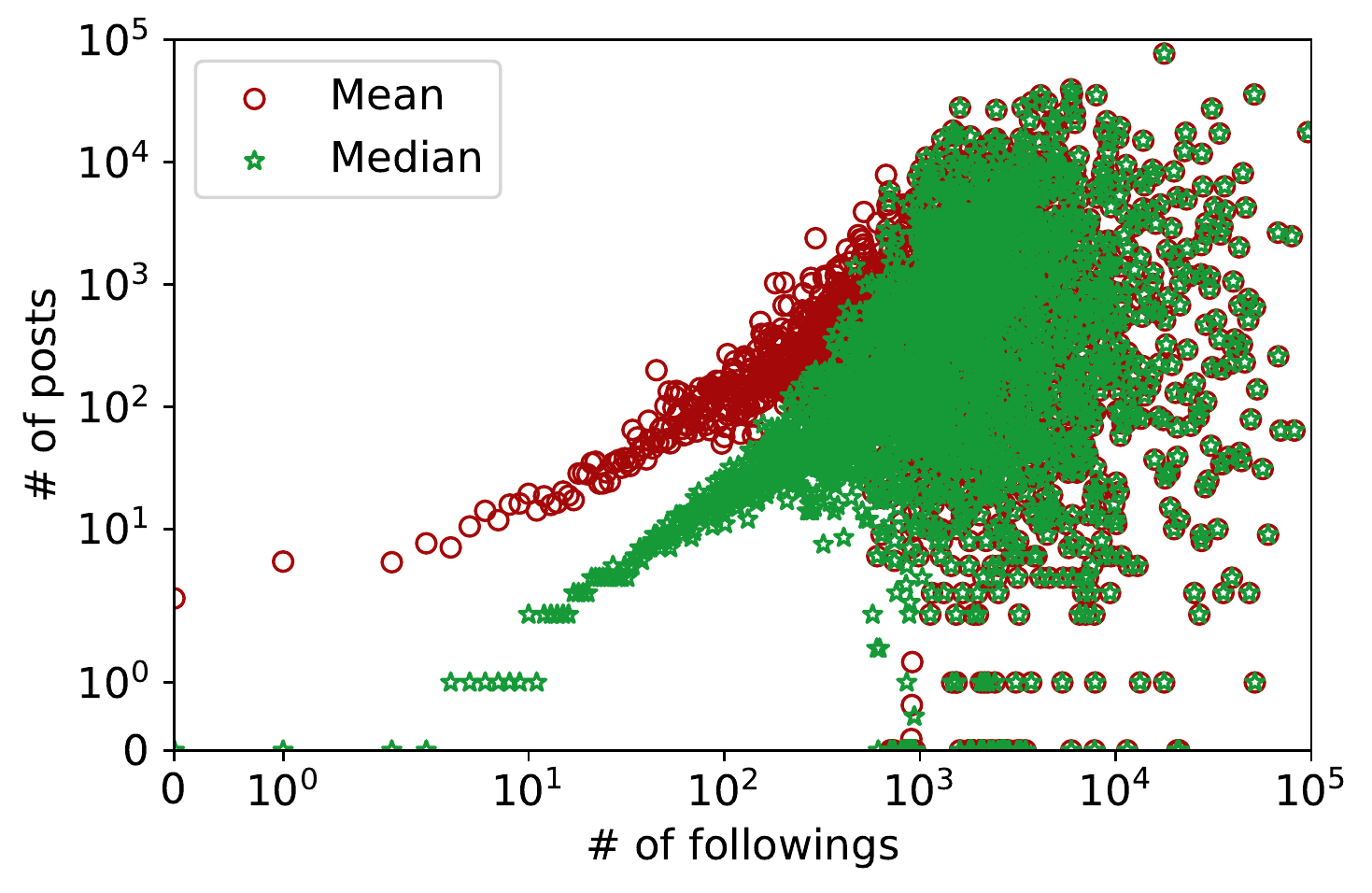}\label{subfig:scatter_following_posts}}
\caption{Followers and Following analysis (a) CDF of number of followers and following (b) number of followers and number of posts and (c) number of following and number of posts. }
\label{fig:followers_followings}
\end{figure*}

\descr{Followers/Followings.} Fig.~\ref{fig:followers_followings} reports our analysis based on the number of followers and followings for each user.
From Fig.~\ref{fig:cdf_followers_following} we observe that in general Gab users have a larger number of followers when compared with following users.
Interestingly, 43\% of users are following zero other users, while only 4\% of users have zero followers.
I.e., although counter-intuitive, most users have more followers than users they follow.
Figs.~\ref{subfig:scatter_followers_posts} and ~\ref{subfig:scatter_following_posts} show the number of followers and following in conjunction with the number of posts for each Gab user. 
We bin the data in log-scale bins and we report the mean and median value for each bin.
We observe that in both cases, that there is a near linear relationship with the number of posts and followers/followings up until around 10 followers/followings.
After this point, we see this relationship diverge, with a substantial number of users with huge numbers of posts, some over 77K.
This demonstrates the extremely heavy tail in terms of content production on Gab, as is typical of most social medial platforms.

\descr{Reciprocity.} From the followers/followings network we find a low level of reciprocity: specifically, only 29.5\% of the node pairs in the network are connected both ways, while the remaining 71.5\% are connected one way.
When compared with the corresponding metric on Twitter~\cite{kwak2010what}, these results highlight that Gab has a larger degree of network reciprocity indicating that the community is more tightly-knit, which is expected when considering that Gab mostly attracts users from the same ideology (i.e., alt-right community).

\subsection{Posts Analysis}
\descr{Basic Statistics.}
First, we note that 63\% of the posts in our dataset are original posts while 37\% are reposts.
Interestingly, only 0.14\%  of the posts are marked as NSFW. 
This is surprising given the fact that one of the reasons that Apple rejected Gab's mobile app is due to the share of NSFW content~\cite{gab_apple_porn}.
From browsing the Gab platform, we also can anecdotally confirm the existence of NSFW posts that are not marked as such, raising questions about how Gab moderates and enforces the use of NSFW tags by users.
When looking a bit closer at their policies, Gab notes that they use a 1964 United States Supreme Court Ruling~\cite{wikipedia_1964} on pornography that provides the famous ``I'll known it when I see it'' test.
In any case, it would seem that Gab's social norms are relatively lenient with respect to what is considered NSFW.

We also look into the languages of the posts, as returned by Gab's API. 
We find that Gab's API does not return a language code for 56\% of posts.
By looking at the dataset, we find that all posts before June 2016 do not have an associated language; possibly indicating that Gab added the language field afterwards.
Nevertheless, we find that the most popular languages are English (40\%), Deutsch (3.3\%), and French (0.14\%); possibly shedding light to Gab's users locations which are mainly the US, the UK, and Germany.

\begin{table}[t]
\centering
\small
\begin{tabular}{lrlr}
\textbf{Domain}      & \multicolumn{1}{l}{\textbf{(\%)}} & \textbf{Domain} & \multicolumn{1}{l}{\textbf{(\%)}} \\ \hline
youtube.com          & 4.22\%                            & zerohedge.com   & 0.53\%                            \\
youtu.be             & 2.67\%                            & twimg.com       & 0.53\%                            \\
twitter.com          & 1.96\%                            & dailycaller.com & 0.49\%                            \\
breitbart.com        & 1.44\%                            & t.co            & 0.47\%                            \\
bit.ly               & 0.82\%                            & ussanews.com    & 0.46\%                            \\
thegatewaypundit.com & 0.74\%                            & dailymail.co.uk & 0.46\%                            \\
kek.gg               & 0.69\%                            & tinyurl.com     & 0.44\%                            \\
imgur.com            & 0.68\%                            & wordpress.com   & 0.43\%                            \\
sli.mg               & 0.61\%                            & foxnews.com     & 0.41\%                            \\
infowars.com         & 0.56\%                            & blogspot.com    & 0.32\%                          \\ \hline 
\end{tabular}
\caption{Top 20 domains in posts and their respective percentage over all posts.}
\label{tbl:top_domains}
\end{table}

\descr{URLs.} Next , we assess the use of URLs in Gab; overall we find 3.5M unique URLs from 81K domains. 
Table~\ref{tbl:top_domains} reports the top 20 domains according to their percentage of inclusion in all posts.
We observe that the most popular domain is YouTube with almost 7\% of all posts, followed by Twitter with 2\%.
Interestingly, we note the extensive use of alternative news sources like Breitbart (1.4\%), The Gateway Pundit (0.7\%), and Infowars (0.5\%), while mainstream news outlets like Fox News (0.4\%) and Daily Mail (0.4\%) are further below.
Also, we note the use of image hosting services like Imgur (0.6\%), \url{sli.mg} (0.6\%), and kek.gg (0.7\%) and URL shorteners like \url{bit.ly} (0.8\%) and \url{tinyurl.com} (0.4\%).
Finally, it is worth mentioning that The Daily Stormer, a well known neo-Nazi web community is five ranks ahead of the most popular mainstream news source, The Hill.

\begin{table}[t]
\centering
\small
\begin{tabular}{lrlr}
\textbf{Hashtag} & \multicolumn{1}{l}{\textbf{(\%)}} & \textbf{Mention} & \multicolumn{1}{l}{\textbf{(\%)}} \\ \hline
MAGA             & \multicolumn{1}{r|}{6.06\%}       & a                & 0.69\%                            \\
GabFam           & \multicolumn{1}{r|}{4.22\%}       & TexasYankee4     & 0.31\%                            \\
Trump            & \multicolumn{1}{r|}{3.01\%}       & Stargirlx        & 0.26\%                            \\
SpeakFreely      & \multicolumn{1}{r|}{2.28\%}       & YouTube          & 0.24\%                            \\
News             & \multicolumn{1}{r|}{2.00\%}       & support          & 0.23\%                            \\
Gab              & \multicolumn{1}{r|}{0.88\%}       & Amy              & 0.22\%                            \\
DrainTheSwamp    & \multicolumn{1}{r|}{0.71\%}       & RaviCrux         & 0.20\%                            \\
AltRight         & \multicolumn{1}{r|}{0.61\%}       & u                & 0.19\%                            \\
Pizzagate        & \multicolumn{1}{r|}{0.57\%}       & BlueGood         & 0.18\%                            \\
Politics         & \multicolumn{1}{r|}{0.53\%}       & HorrorQueen      & 0.17\%                            \\
PresidentTrump   & \multicolumn{1}{r|}{0.47\%}       & Sockalexis       & 0.17\%                            \\
FakeNews         & \multicolumn{1}{r|}{0.41\%}       & Don              & 0.17\%                            \\
BritFam          & \multicolumn{1}{r|}{0.37\%}       & BrittPettibone   & 0.16\%                            \\
2A               & \multicolumn{1}{r|}{0.35\%}       & TukkRivers       & 0.15\%                            \\
maga             & \multicolumn{1}{r|}{0.32\%}       & CurryPanda       & 0.15\%                            \\
NewGabber        & \multicolumn{1}{r|}{0.28\%}       & Gee              & 0.15\%                            \\
CanFam           & \multicolumn{1}{r|}{0.27\%}       & e                & 0.14\%                            \\
BanIslam         & \multicolumn{1}{r|}{0.25\%}       & careyetta        & 0.14\%                            \\
MSM              & \multicolumn{1}{r|}{0.22\%}       & PrisonPlanet     & 0.14\%                            \\
1A               & \multicolumn{1}{r|}{0.21\%}       & JoshC            & 0.12\%                            \\ \hline
\end{tabular}
\caption{Top 20 hashtags and mentions found in Gab. We report their percentage over all posts.}
\label{tbl:top_hashtags_mentions}
\end{table}

\descr{Hashtags \& Mentions}
As discussed in Section~\ref{sec:gab}, Gab supports the use of hashtags and mentions similar to Twitter. 
Table~\ref{tbl:top_hashtags_mentions} reports the top 20 hashtags/mentions that we find in our dataset.
We observe that the majority of the hashtags are used in posts about Trump, news, and politics.
We note that among the top hashtags are ``AltRight'', indicating that Gab users are followers of the alt-right movement or they discuss topics related to the alt-right; ``Pizzagate'', which denotes discussions around the notorious conspiracy theory~\cite{bbc_4chan_pizzagate}; and ``BanIslam'', which indicate that Gab users are sharing their islamophobic views.
It is also worth noting the use of hashtags for the dissemination of popular memes, like the Drain the Swamp meme that is popular among Trump's supporters~\cite{vox_memes_trump}.
When looking at the most popular users that get mentioned, we find popular users related to the Gab platform like Andrew Torba (Gab's CEO with username @a).

We also note users that are popular with respect to mentions, but do \emph{not} appear in Table~\ref{tbl:top_20_users}'s lists of popular users.
For example, Amy is an account purporting to be Andrew Torba's mother. 
The user Stargirlx, who we note changed usernames three times during our collection period, appears to be an account presenting itself as a millennial ``GenZ'' young woman.
Interestingly, it seems that Amy and Stargirlx have been organizing Gab ``chats,'' which are private groups of users, for 18 to 29 year olds to discuss politics; possibly indicating efforts to recruit millennials to the alt-right community.

\begin{figure*}[t!]
\centering
\subfigure[Date]{\includegraphics[width=0.32\textwidth]{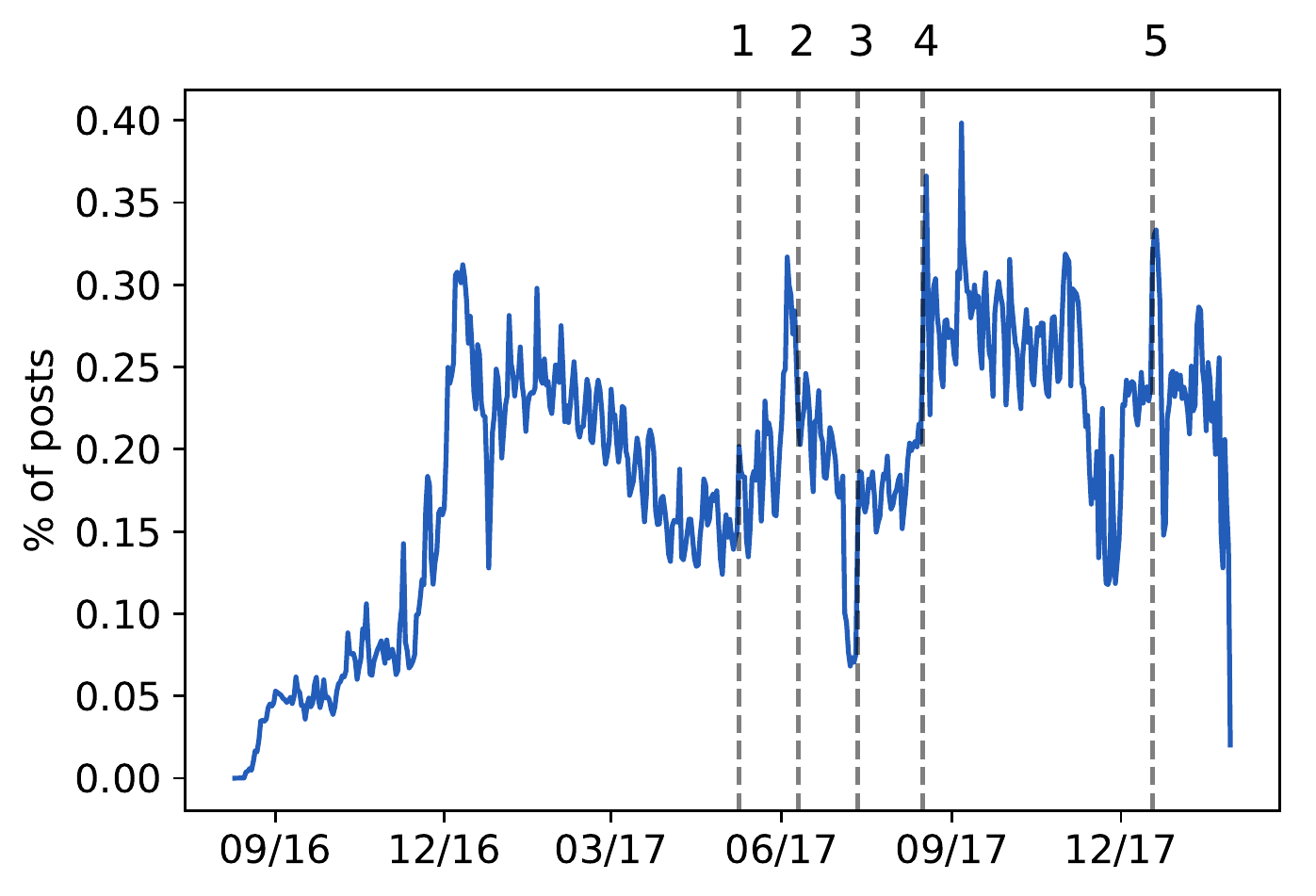}\label{subfig:counts_day}}
\subfigure[Hour of Day]{\includegraphics[width=0.32\textwidth]{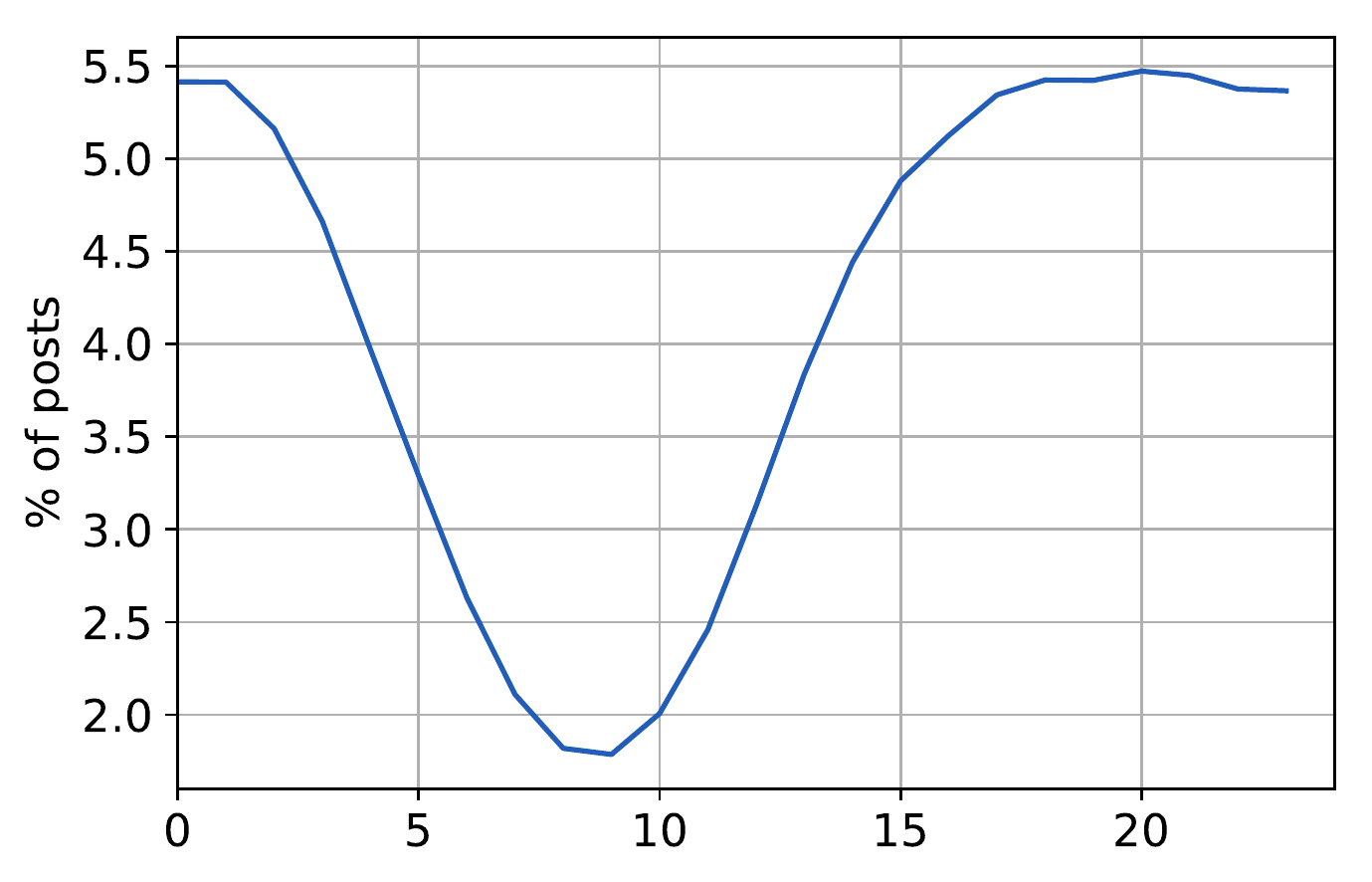}\label{subfig:counts_hour_day}}
\subfigure[Hour of Week]{\includegraphics[width=0.32\textwidth]{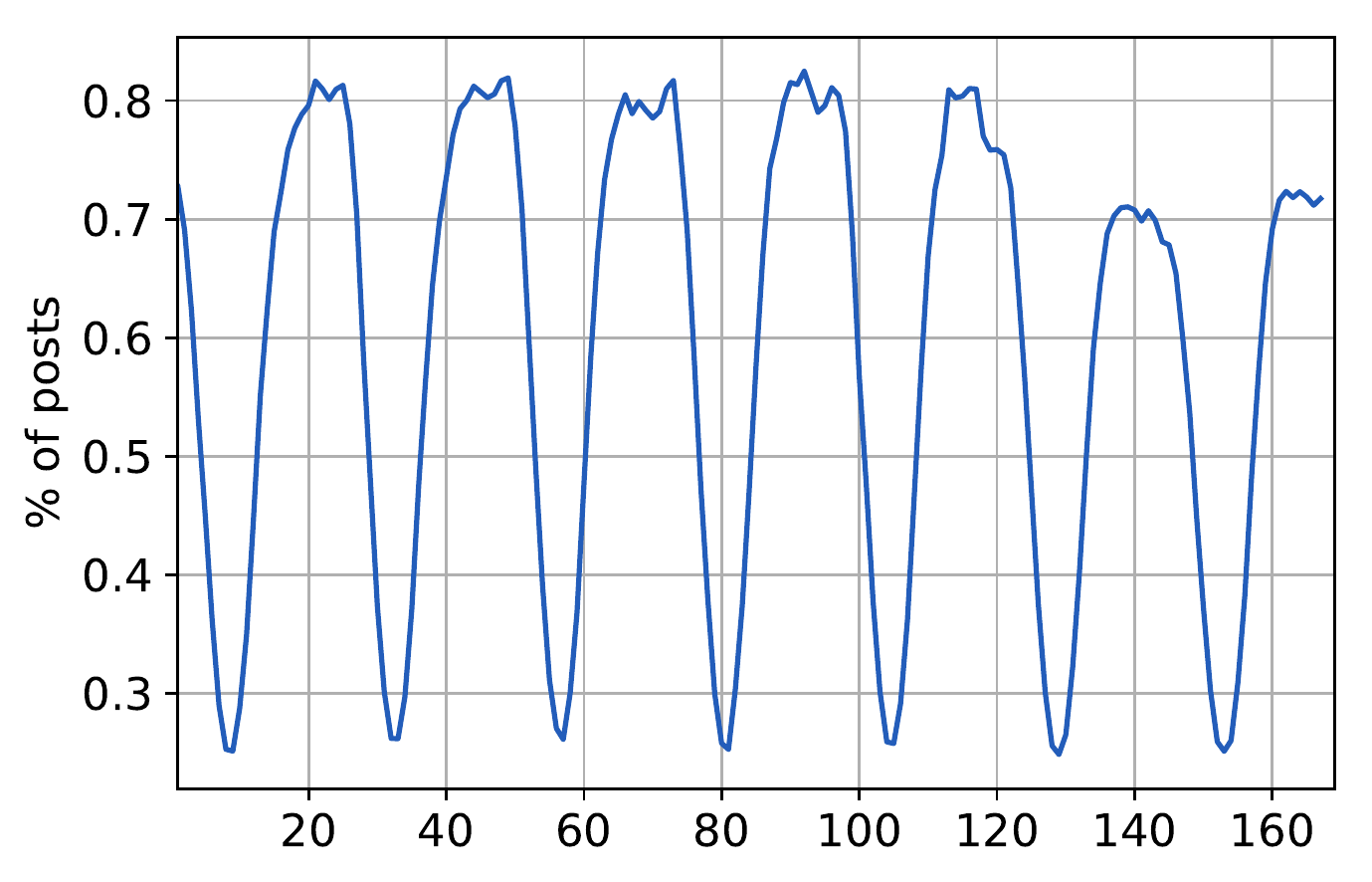}\label{subfig:counts_hour_week}}
\caption{Temporal analysis of the Gab posts (a) each day; (b) based  on hour of day and (c) based on hour of week.}
\label{fig:temporal_analysis}
\end{figure*}

\descr{Categories \& Topics.} As discussed in Section~\ref{sec:gab} gabs may be part of a topic or category.
By analyzing the data, we find that this happens for 12\% and 42\% of the posts for topics and categories, respectively.
Table~\ref{tbl:top_categories_topics} reports the percentage of posts for each category as well as for the top 15 topics.
For topics, we observe that the most popular are general ``Ask Me Anything'' (AMA) topics like Deutsch (2.29\%, for German users), BritFam (0.73\%, for British users), and Introduce Yourself (0.59\%). 
Furthermore, other popular topics include world events and news like International News (0.59\%), Las Vegas shooting (0.27\%), and conspiracy theories like Seth Rich's Murder (0.11\%). 
When looking at the top categories we find that by far the most popular categories are News (15.91\%) and Politics (10.30\%).
Other popular categories include AMA 4.46\%), Humor (3.50\%), and Technology (1.44\%).

These findings highlight that Gab is heavily used for the dissemination and discussion of world events and news.
Therefore, its role and influence on the Web's information ecosystem should be assessed in the near future.
Also, this categorization of posts can be of great importance for the research community as it provides labeled ground truth about discussions around a particular topic and category.

\begin{table}[]
\centering
\small
\begin{tabular}{lrlr}
\textbf{Topic}           & \multicolumn{1}{l}{\textbf{(\%)}} & \textbf{Category} & \multicolumn{1}{l}{\textbf{(\%)}} \\ \hline
Deutsch                  & \multicolumn{1}{r|}{2.29\%}       & News              & 15.91\%                           \\
BritFam                  & \multicolumn{1}{r|}{0.73\%}       & Politics          & 10.30\%                           \\
Introduce Yourself       & \multicolumn{1}{r|}{0.59\%}       & AMA               & 4.46\%                            \\
International News       & \multicolumn{1}{r|}{0.19\%}       & Humor             & 3.50\%                            \\
DACA                     & \multicolumn{1}{r|}{0.17\%}       & Technology        & 1.44\%                            \\
Las Vegas Terror Attack  & \multicolumn{1}{r|}{0.16\%}       & Philosophy        & 1.06\%                            \\
Hurricane Harvey         & \multicolumn{1}{r|}{0.16\%}       & Entertainment     & 1.01\%                            \\
Gab Polls                & \multicolumn{1}{r|}{0.13\%}       & Art               & 0.72\%                            \\
London                   & \multicolumn{1}{r|}{0.12\%}       & Faith             & 0.69\%                            \\
2017 Meme Year in Review & \multicolumn{1}{r|}{0.12\%}       & Science           & 0.56\%                            \\
Twitter Purge            & \multicolumn{1}{r|}{0.12\%}       & Music             & 0.52\%                            \\
Seth Rich                & \multicolumn{1}{r|}{0.11\%}       & Sports            & 0.39\%                            \\
Memes                    & \multicolumn{1}{r|}{0.11\%}       & Photography       & 0.37\%                            \\
Vegas Shooting           & \multicolumn{1}{r|}{0.11\%}       & Finance           & 0.31\%                            \\
Judge Roy Moore          & \multicolumn{1}{r|}{0.09\%}       & Cuisine           & 0.16\%                            \\ \hline
\end{tabular}
\caption{Top 15 categories and topics found in the Gab dataset}
\label{tbl:top_categories_topics}
\end{table}

\descr{Hate speech assessment.} As previously discussed, Gab was openly accused of allowing the dissemination of hate speech. 
In fact, Google removed Gab's mobile app from its Play Store because it violates their hate speech policy~\cite{gab_hate_speech}.
Due to this, we aim to assess the extent of hate speech in our dataset.
Using the modified Hatebase~\cite{hatebase} dictionary used by the authors of~\cite{hine2017kek}, we find that 5.4\% of all Gab posts include a hate word.
In comparison, Gab has 2.4 times the rate of hate words when compared to Twitter, but less than halve the rate of hate words compared to 4chan's Politically Incorrect board (/pol/)~\cite{hine2017kek}.
These findings indicate that Gab resides on the border of mainstream social networks like Twitter and fringe Web communities like 4chan's Politically Incorrect (/pol/) board.

\descr{Temporal Analysis.} Finally, we study the posting behavior of Gab users from a temporal point of view. Fig.~\ref{fig:temporal_analysis} shows the distribution of the Gab posts in our dataset according to each day of our dataset, as well as per hour of day and week (in UTC).
We observe that the general trend is that the number of Gab's posts increase over time (Fig.~\ref{subfig:counts_day}); this indicates an increase in Gab's popularity.
Furthermore, we note that Gab users posts most of their gabs during the afternoon and late night (after 3 PM UTC) while they rarely post during the morning hours (Fig.~\ref{subfig:counts_hour_day}).
Also, the aforementioned posting behavior follow a diurnal weekly pattern as we show in Fig.~\ref{subfig:counts_hour_week}.

To isolate significant days in the time series in Fig.~\ref{subfig:counts_day}, we perform a \emph{changepoint analysis} using the Pruned Exact Linear Time (PELT) method~\cite{changepoint}.
First, we use our knowledge of the weekly variation in average post numbers from Fig.~\ref{subfig:counts_hour_week} to subtract from our timeseries the mean number of posts for each day.
This leaves us with a mean-zero timeseries of the deviation of the number of posts per day from the daily average.
We assume that this timeseries is drawn from a normal distribution, with mean and variance that can change at a discrete number of changepoints.
We then use the PELT algorithm to maximize the log-likelihood function for the mean(s) and variance(s) of this distribution, with a penalty for the number of changepoints.
By ramping down the penalty function, we produce a ranking of the changepoints.

Examining current events around these changepoints provides insight into they dynamics that drive Gab behavior.
First, we note that there is a general increase in activity up to the Trump inauguration, at which point activity begins to decline.
When looking later down the timeline, we see an increase in activity after the changepoint marked \textbf{1} in Fig.~\ref{subfig:counts_day}.
Changepoint \textbf{1} coincides with James Comey's firing from the FBI, and the relative acceleration of the Trump-Russian collusion probe~\cite{james_comey}.

The next changepoint (\textbf{2}) coincides with the so-called ``March Against Sharia''~\cite{sharia} organized by the alt-right, with the event marked \textbf{4} corresponding to Trump's ``blame on both sides'' response to violence at the Unite the Right Rally in Charlottesville~\cite{charlotesville_blame}.
Similarly, we see a meaningful response to Twitter's banning of abusive users~\cite{twitter_purge} marked as changepoint \textbf{5}.

Changepoint \textbf{3}, occurring on July 12, 2017 is of particular interest, since it is the most extreme \emph{reduction} in activity recognized as a changepoint.
From what we can tell, this is a reaction to Donald Trump Jr. releasing emails that seemingly evidenced his meeting with a Russian lawyer to receive compromising intelligence on Hillary Clinton's campaign~\cite{donald_trump_junior}.
I.e., the disclosure of evidence of collusion with Russia corresponded to the single largest drop in posting activity on Gab.

\section{Conclusion} \label{sec:conclusion}
In this work. we have provided the first characterization of a new social network called Gab.
We analyzed 22M posts from 336K users, finding that Gab attracts the interest of users ranging from alt-right supporters and conspiracy theorists to trolls.
We showed that Gab is extensively used for the discussion of news, world events, and politics-related topics, further motivating the need take it into account when studying information cascades on the Web.
By looking at the posts for hate words, we also found that 5.4\% of the posts include hate words.
Finally, using changepoint analysis, we highlighted how Gab reacts very strongly to real-world events focused around white nationalism and support of Donald Trump.

\smallskip\noindent\textbf{Acknowledgments.} This project has received funding from the European Union's Horizon 2020 Research and Innovation program under the Marie Sk\l{}odowska-Curie ENCASE project (Grant Agreement No. 691025).
The work reflects only the authors' views; the Agency and the Commission are not responsible for any use that may be made of the information it contains.

\bibliographystyle{abbrv}

\end{document}